\documentclass[]{article}
\usepackage{spconf,amsmath,graphicx,amssymb,delarray,subfigure,verbatim,booktabs}
\usepackage{diagbox, makecell}
\usepackage{multirow}
\usepackage[english]{babel}
\usepackage{lipsum}
\usepackage{bbding}
\usepackage{setspace}

\title{Dual-branch Attention-In-Attention Transformer for single-channel speech enhancement}
\name{Guochen Yu$^{\star \dagger}$\thanks{This work was supported by the National Natural Science Foundation  of China [Grant 61631016] and National Key R\&D Program of China [Grant No. SQ2020YFF0426386]. Hui Wang is the corresponding author.}, Andong Li$^{\dagger}$, Chengshi Zheng$^{\dagger}$, Yinuo Guo$^{\ast}$, Yutian Wang$^{\star }$, Hui Wang$^{\star}$ }
\address{$^{\star}$ State Key Laboratory of Media Convergence and Communication, Communication University \\of China, Beijing, China\\
	$^{\dagger}$  Institute of Acoustics, Chinese Academy of Sciences, Beijing, China\\
	$^{\ast}$  Bytedance, Beijing, China\\
	\{yuguochen, wangyutian, hwang\}@cuc.edu.cn, \{liandong, cszheng\}@mail.ioa.ac.cn }

\begin{document}
	\ninept
	\maketitle
	\begin{abstract}
		\vspace{-2mm}
		Curriculum learning begins to thrive in the speech enhancement area, which decouples the original spectrum estimation task into multiple easier sub-tasks to achieve better performance. Motivated by that, we propose a dual-branch attention-in-attention transformer dubbed DB-AIAT to handle both coarse- and fine-grained regions of the spectrum in parallel. From a complementary perspective, a magnitude masking branch is proposed to coarsely estimate the overall magnitude spectrum, and simultaneously a complex refining branch is elaborately designed to compensate for the missing spectral details and implicitly derive phase information. Within each branch, we propose a novel attention-in-attention transformer-based module to replace the conventional RNNs and temporal convolutional networks for temporal sequence modeling. Specifically, the proposed attention-in-attention transformer consists of adaptive temporal-frequency attention transformer blocks and an adaptive hierarchical attention module, aiming to capture long-term temporal-frequency dependencies and further aggregate global hierarchical contextual information.  Experimental results on Voice Bank + DEMAND demonstrate that DB-AIAT yields state-of-the-art performance (\emph{e.g.}, 3.31 PESQ, 95.6\% STOI and 10.79dB SSNR) over previous advanced systems with a relatively small model size (2.81M).
		
		
	\end{abstract}
	
	\begin{keywords}
		Speech enhancement, dual-branch, attention-in-attention, transformer
	\end{keywords}
	
	\vspace{-5mm}
	\section{Introduction}
	\vspace{-3mm}
	In real acoustic scenarios, various types of environmental interference may severely degrade the performance of telecommunication and hearing aids. Monaural speech enhancement (SE) technique aims at recovering clean speech from its noise-corrupted mixture to improve speech quality and/or intelligibility when only one channel recording is available~{\cite{loizou2013speech}}. Recently, deep neural networks (DNNs) have ignited the development of SE algorithms for their more powerful capability in dealing with non-stationary noise than conventional statistical signal-processing based approaches~{\cite{wang2018supervised}}.
	
	In a typical supervised SE paradigm, DNNs are usually leveraged to estimate the mask functions or directly predict the magnitude spectra of clean speech in the time-frequency (T-F) domain~{\cite{wang2014training, xu2014regression}}, where the noisy phase is unchanged and involved in waveform reconstruction. Recently, the importance of phase has been fully illustrated in improving the speech perceptual quality, especially under low signal-to-noise ratio (SNR) conditions~{\cite{paliwal2011importance}}. In this regard, more recent approaches attempt to recover the phase information either explicitly or implicitly~{\cite{tan2019learning, li2021two, hu2020dccrn, choi2019phase, wang2021tstnn, defossez2020real}}. For the first class, the network is employed to estimate the complex ratio masks (CRMs) or the real and imaginary (RI) spectra, which facilitate both magnitude and phase information recovery simultaneously in the T-F domain. For the latter, the time-domain waveform is directly regenerated, which diverts around the phase estimation process. More recently, decoupling-style phase-aware methods are proposed, where the original complex-spectrum estimation problem is decomposed into two sub-stages~{\cite{li2021two, li2021simultaneous, li2021glance}}. Specifically, only the magnitude estimation is involved in the first stage, followed by the spectrum refinement with residual learning in the later stage. In this way, the optimization \emph{w.r.t.} magnitude and phase can be effectively decoupled, which alleviates the implicit compensation effect between two targets~{\cite{wang2021compensation}}.
	
	In this paper, we propose a dual-branch SE structure dubbed DB-AIAT, to explore the complex spectrum recovery from the complementary perspective. Specifically, two core branches are elaborately designed in parallel, namely a magnitude masking branch (MMB) and a complex refining branch (CRB). In MMB, we seek to construct the filtering system which only applies to the magnitude domain. In this branch, most of the noise can be effectively suppressed. In CRB, it is established as the decorating system to compensate for the lost spectral details and phase mismatch effect. Two branches work collaboratively to facilitate the overall spectrum recovery. Besides, despite temporal convolutional networks (TCNs)~{\cite{pandey2019tcnn}} and LSTM layers are widely adopted for long-range sequence modeling in the SE area, they still lack sufficient capacity to capture the global context information~{\cite{wang2021tstnn, chen2020dual}}. In addition, they usually only apply in the time axis, which neglects the correlations among different frequency sub-bands. To this end, we propose an attention-in-attention transformer (AIAT) to funnel the global sequence modeling process, which captures long-range dependencies along both time and frequency axes, and meanwhile, aggregates global hierarchical contextual information. Experimental results on Voice Bank + DEMAND dataset show that DB-AIAT achieves remarkable results and consistently outperforms state-of-the-art baselines with a relatively light model size.
	
	The remainder of the paper is organized as follows. In Section~{\ref{Sec2}}, the proposed framework is described in detail. The experimental setup is presented in Section~{\ref{Sec3}}, while Section~{\ref{Sec4}} gives the results and analysis. Finally, some conclusions are drawn in Section~{\ref{Sec5}}.
	\begin{figure*}[t]
		\centering
		\centerline{\includegraphics[width=1.87\columnwidth]{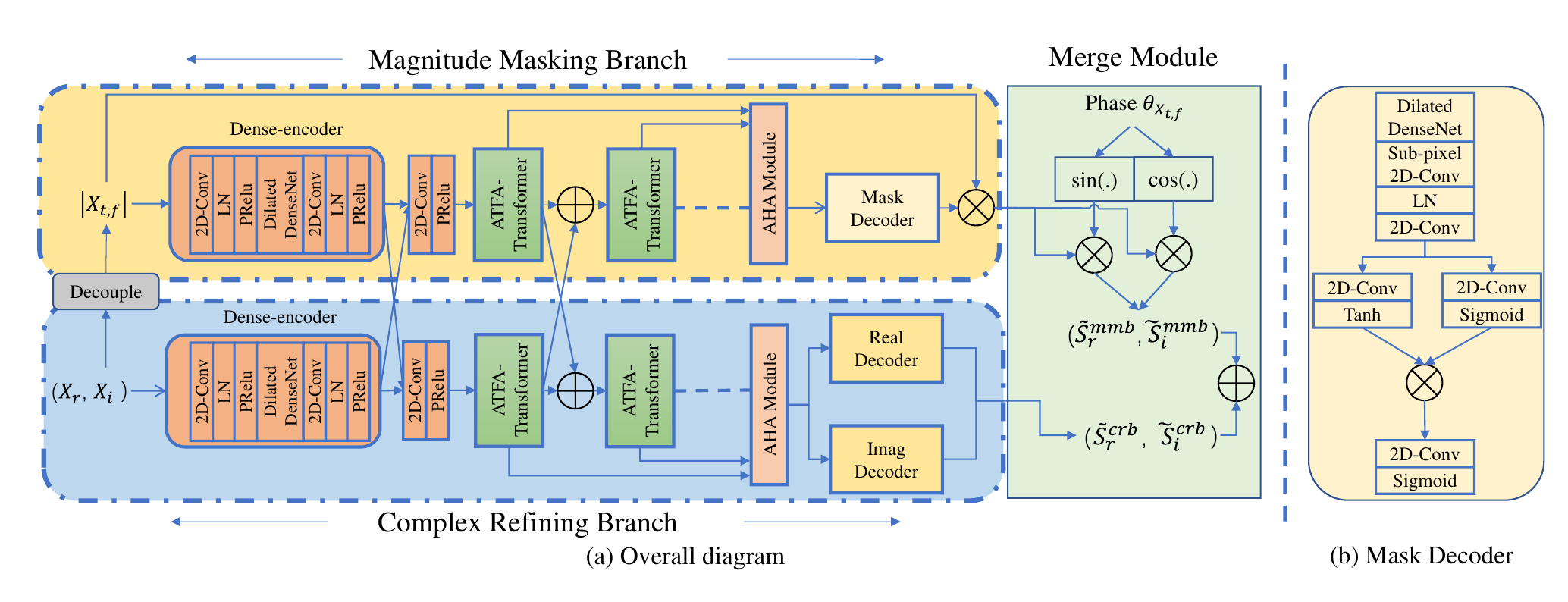}}
		\vspace{-0.6cm}
		\caption{The diagram of the proposed DB-AIAT. (a) The overall diagram of the proposed system. (b) The detailed architecture of mask decoder.}
		\label{fig:diagram-system}
		\vspace{-0.2cm}
	\end{figure*}
	\vspace{-4mm}
	\section{Methodlogy\label{Section2}}
	\label{Sec2}
	\vspace{-3mm}
	\subsection{ Dual-branch magnitude masking and complex refining \label{Section23}}
	\vspace{-2mm}
	The overall diagram of the proposed system is illustrated in Fig.~{\ref{fig:diagram-system}}. It is mainly comprised of two branches, namely a magnitude spectrum masking branch (MMB) and a complex spectrum refining branch (CRB), which aim at collaboratively estimating the magnitude and phase information of clean speech in parallel.    
	To be specific, in the MMB path, the input is the magnitude of the noisy spectrum, and the network estimates the magnitude mask $M^{mmb}$ to coarsely recover the magnitude of the target speech, \emph{i.e.}, $\lvert\widetilde{S}^{mmb}\rvert$. Subsequently, the coarsely estimated spectral magnitude is coupled with the noisy phase to derive the coarse-denoised complex spectrum.
	
	As the complement, CRB receives noisy RI component $\left\{ X_{r}, X_{i}\right\}$ as the input and focuses on the fine-grained spectral structures which may be lost in the MMB path and further suppressing the residual noise components. Note that we only estimate the residual details instead of explicitly estimating the whole complex spectrum, which alleviates the overall burden of the network. The alternate interconnections are adopted to exchange information between the two branches, enabling better feature representation. Finally, we sum the coarse-denoised complex spectrum and the fine-grained complex spectral details together to reconstruct the clean complex spectrum. In a nutshell, the whole procedure can be formulated as:
	\begin{gather}
		\label{eqn1}
		\lvert\widetilde{S}^{mmb}\rvert = \lvert{X_{t,f}}\rvert \otimes M^{mmb},\\
		\widetilde{S}^{mmb}_{r} = \vert\widetilde{S}^{mmb}\rvert \otimes \cos\left( \theta_{X}\right),\\
		\widetilde{S}^{mmb}_{i} = \lvert\widetilde{S}^{mmb}\rvert \otimes \sin\left( \theta_{X} \right), \\
		\widetilde{S}_{r} = \widetilde{S}^{mmb}_{r} + \widetilde{S}^{crb}_{r}, \\
		\widetilde{S}_{i} = \widetilde{S}^{mmb}_{i} + \widetilde{S}^{crb}_{i}
		\vspace{-0.2cm}	
	\end{gather}
	where $\left\{ \widetilde{S}^{crb}_{r}, \widetilde{S}^{crb}_{i}  \right\}$ denote the output RI components of CRB and $\left\{\widetilde{S}_{r}, \widetilde{S}_{i}\right\}$ denote the final merged clean RI components. Tilde denotes the estimated variable.  $\theta_{X}$ denotes the noisy phase and $\otimes$ is the element-wise multiplication operator. The input features of MMB and CRB are denoted as $\lvert{X_{t,f}}\rvert \in \mathbb{R}^{ T\times F\times 1}$ and $Cat(X_r, X_i)\in \mathbb{R}^{T\times F\times 2}$, respectively. Here $T$ is the number of frames and $F$ is the number of frequency bins.
	As shown in Fig.~{\ref{fig:diagram-system}}, MMB consists of a densely-connected convolutional encoder, an AIAT and a mask decoder. Analogously, CRB is composed of a dense-encoder, an AIAT and two complex decoders.
	\vspace{-0.4cm} 
	\subsection{Densely convolutional encoder \label{Section21}}
	\vspace{-0.2cm}
	The dense-encoder in each branch is composed of two 2-D convolutional layers, followed by layer normalization (LN) and parametric ReLU (PReLU) activation. A densenet with four dilated convolutional layers is employed between the above 2-D convolutional layers, in which the dilation rates are $\left\{1,2,4,8\right\}$. The output channel of the first 2-D convolutional layer is set to 64 and keeps unaltered, with kernel size and stride being (1, 1), while the second 2-D convolutional layer halves the dimension of the frequency axis, with kernel size and stride being (1, 3) and (1, 2), respectively.
	\vspace{-0.0cm}
	\begin{figure}[t]
		\centering
		\centerline{\includegraphics[width=0.87\columnwidth]{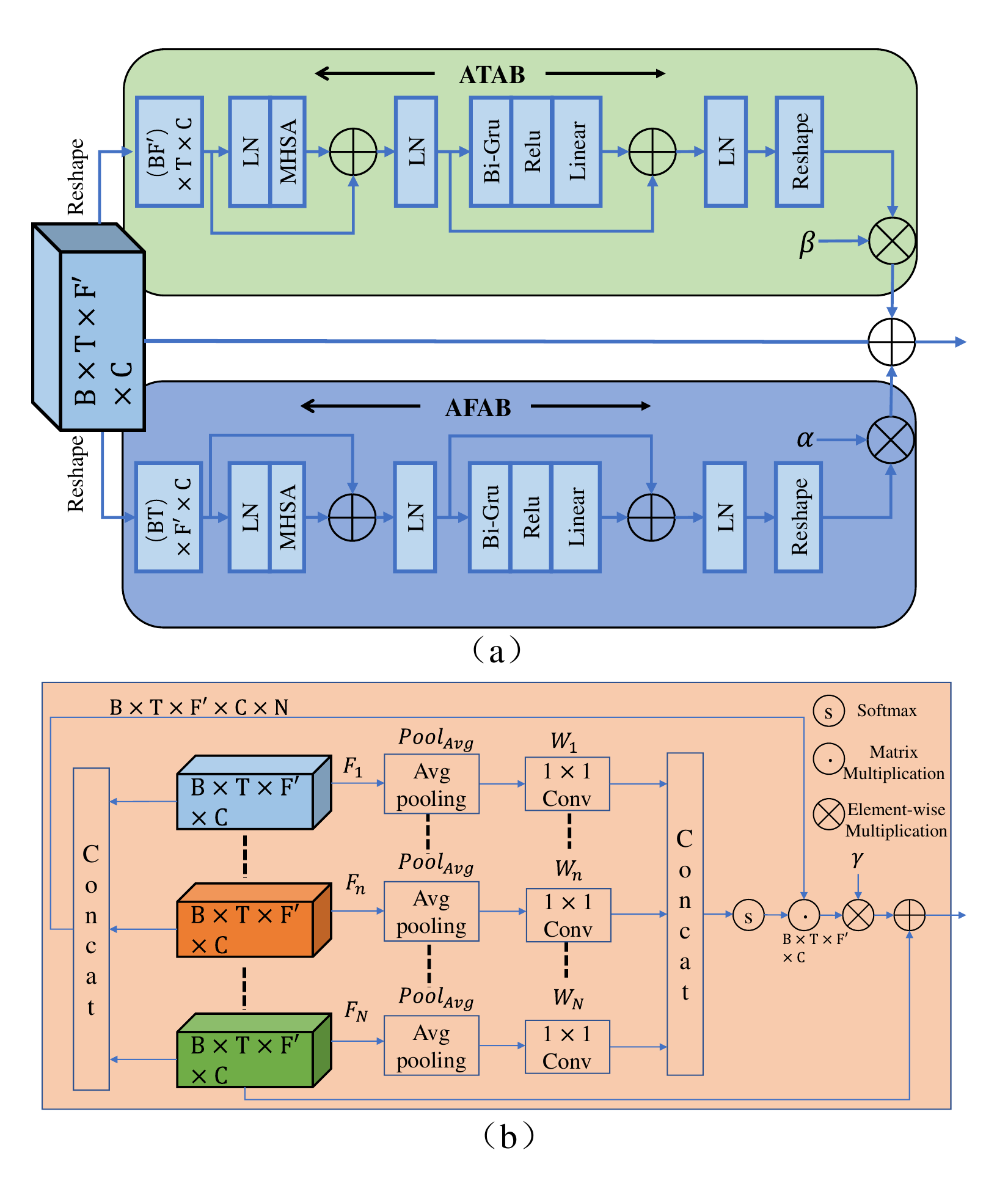}}
		\vspace{-0.5cm}
		\caption{ (a) The diagram of ATFAT blocks. (b) The diagram of the AHA module.}
		\vspace{-0.5cm}
		\label{fig:ATFA}	
	\end{figure}
	\vspace{-0.5cm}  	
	\subsection{ Attention-in-attention transformer \label{Section22}}
	\vspace{-2mm} 
	The proposed AIAT module consists of four adaptive time-frequency attention transformer-based (ATFAT) modules and an adaptive hierarchical attention (AHA) module, as illustrated in Fig.~{\ref{fig:ATFA}}. Each ATFAT can strengthen the long-range temporal-frequency dependencies with relatively low computational cost and the AHA module can aggregate different intermediate features to capture multi-scale contextual information by adaptive weights, as pointed out in~{\cite{yu2021cyclegan}}. Before feeding the compressed output into AIAT, we concatenate the output of two dense-encoders and a 2-D convolution layer with kernel size (1, 1) is used to halve the number of channels (i.e., from 128 to 64), followed by PReLU activation. As shown in Fig.~{\ref{fig:ATFA}}(a), ATFAT is composed of an adaptive temporal attention branch (ATAB) and an adaptive frequency attention branch (AFAB), which can capture long-term dependencies along temporal and frequency axes with two adaptive weights $\alpha$ and $\beta$. In each branch, an improved transformer~{\cite{wang2021tstnn}} is employed, which is comprised of a multi-head self-attention (MHSA) module and a GRU-based feed-forward network, followed by residual connections and LN. The feed-forward network employs a bi-directional GRU (Bi-GRU) to replace the first fully connected layer in the traditional transformer and yields better performance~{\cite{wang2021tstnn}}. After each ATFAT module, we employ PReLU and a 2-D convolution with kernel size (1, 1) to retain the channel number.
	
	The AHA module aims at integrating different hierarchical feature maps given all the ATFAT modules' outputs $F = \left \{ F_n \right \}^{N}_{n=1}, F_n \in \mathbb{R}^{B\times T\times F'\times C}$, where $B$ denotes the batch size, $F'$ denotes the halved frequency dimension and $N$ is the number of ATFAT (set $N$=4 in this paper). We first employ an average pooling layer $Pool_{Avg}$ and a $1\times1$ convolutional layer $W_{n}$ to squeeze the output feature of each ATFAT into a global representation: $P_n^h=Pool_{Avg}(F_n)\ast W_{n}\in R^{B\times 1\times 1\times 1}$. Then we concatenate all the pooled outputs as $P^h \in \mathbb{R}^{B\times 1\times 1\times N\times 1}$, which is subsequently fed into a softmax function to derive the hierarchical attention map $W^h \in \mathbb{R}^{B\times 1\times 1\times N\times 1}$. After that we concatenate all the intermediate outputs of ATFAT (\emph{i.e.}, $\left \{ F_n \right \}^{N}_{n=1}$) to obtain a global feature map $F^h \in \mathbb{R}^{B\times T\times F'\times C\times N}$. Subsequently, we obtain the multi-scale contextual information by performing a matrix multiplication between $F^h$ and $W^h$, which can be calculated as:
	\begin{equation}
		\vspace{-2mm}
		\begin{gathered}		
			G_{N} = W^hF^h
			=\sum_{n=1}^{N} \frac{\textup{exp}(Pool_{Avg}(F_n)\ast W_n)}{\sum_{n=1}^{N}\textup{exp}(Pool_{Avg}(F_n)\ast W_n)}F_{n} ,
		\end{gathered}
	\end{equation}
	where $F_n$ denotes the $n^{th}$ intermediate output of ATFAT and $G_{N}    \in \mathbb{R}^{B\times T\times F'\times C}$ denotes the global contextual feature map. Finally, we perform an element-wise sum operation between the last ATFAT module's output feature map $F_{N}$ and the global contextual feature map $G_{N}$ to obtain the final output, \emph{i.e.}, $Out_{AHA}\in \mathbb{R}^{T\times F'\times C}$:
	\begin{equation}\\
		\vspace{-2mm}
		\begin{gathered}
			Out_{AHA} = F_{N} + \gamma G_{N} ,\\ 
		\end{gathered}
	\end{equation}
	where $\gamma$ is a learnable scalar coefficient and initialized to 0. This adaptive learning weight automatically learns to assign a suitable value to merge global contextual information. 
	\vspace{-0.4cm} 
	\subsection{Mask/Complex decoder \label{Section26}}
	\vspace{-0.2cm}
	The mask decoder consists of a dilated dense block the same as in dense-encoder, a sub-pixel 2-D convolution module, and a dual-path mask module. The sub-pixel convolution layer is used to upsample the compressed features, which demonstrates to be effective in high-resolution image generation~{\cite{shi2016real}}. Then, a dual-path mask module is operated to obtain the magnitude gain function by a 2-D convolution and a dual-path tanh/sigmoid nonlinearity operation, followed by a 2-D convolution and a sigmoid activation function. The final coarse-denoised spectral magnitude is obtained by the element-wise multiplication between the input spectral magnitude and the estimated gain function. In the CRB, real and imaginary decoders are applied to reconstruct both RI components in parallel, which are also composed of a dilated dense block and a sub-pixel 2-D convolution. The sub-pixel 2-D convolution in the mask/complex decoders sets the upsampling factor to 2, with kernel size set to (1, 3).    
	\vspace{-3mm}
	\subsection{Loss function \label{Section24}}
	\vspace{-2mm} 
	The loss function of the proposed two-branch model is calculated by the RI components and the magnitude of the estimated spectrum, which can be expressed as:
	\begin{gather}	
		\mathcal{L}^{Mag}=\left \| \sqrt{\left |\widetilde{S}_r  \right |^2+\left |\widetilde{S}_i  \right |^2 } -  \sqrt{\left |S_r  \right |^2+\left |S_i  \right |^2    } \right \|_{F}^2,\\
		\mathcal{L}^{RI}=\left \|\widetilde{S}_r-S_r \right \|_{F}^2 +\left \|\widetilde{S}_i-S_i \right \|_{F}^2,\\
		\mathcal{L}_{Full}=\mu \mathcal{L}^{RI}+(1-\mu ) \mathcal{L}^{Mag} ,
	\end{gather}
	where $\mathcal{L}^{Mag}$ and $\mathcal{L}^{RI}$ denote the loss function toward magnitude and RI, respectively. Here, $\left\{ \widetilde{S}_r, \widetilde{S}_i  \right\}$ represent the RI components of the estimated speech spectrum, while $\left\{S_r, S_i\right\}$ represent the target RI components of the clean speech spectrum. With the internal trial, we empirically set $\mu= 0.5$ in all the following experiments.
	\vspace{-3mm}
	\section{Experiments\label{Section3}}
	\label{Sec3}
	\vspace{-3mm} 
	\subsection{Datasets\label{Section31}}
	\vspace{-2mm} 
	The dataset used in this work is publicly available as proposed in~{\cite{valentini2016investigating}}, which is a selection of Voice Bank corpus~{\cite{veaux2013voice}} with 28 speakers for training and another 2 unseen speakers for testing. The training set includes 11,572 noisy-clean pairs, while the test set contains 824 pairs. For training, the audio samples are mixed with one of the 10 noise types, (\emph{i.e.}, two artificial (babble and speech shaped) and eight real recording noise processes taken from the DEMAND database~{\cite{thiemann2013diverse}}) at four SNRs, \emph{i.e.}, $\left\{0\rm{dB},5\rm{dB},10\rm{dB},15\rm{dB}\right\}$. The test utterances are created with 5 unseen noise taken from the Demand database at SNRs of $\left\{2.5\rm{dB}, 7.5\rm{dB}, 12.5\rm{dB}, 17.5\rm{dB}\right\}$.
	\vspace{-4mm}
	\subsection{Implementation setup\label{Section32}}
	\vspace{-1mm}  
	All the utterances are resampled at 16 kHz and chunked to 3 seconds. The Hanning window of length 20 ms is selected, with 50\% overlap between consecutive frames. The 320-point STFT is utilized and 161-dimension spectral features are obtained. Due to the efficacy of the compressed spectrum in dereverberation and denoising task~{\cite{li2021simultaneous, li2021importance}}, we conduct the power compression toward the magnitude while remaining the phase unaltered, and the optimal compression coefficient is set to 0.5, \emph{i.e.}, $Cat\left(\lvert X \rvert^{0.5}\cos\left({\theta_{X}}\right), \lvert X \rvert^{0.5}\sin\left({\theta_{X}}\right)\right)$ as input, $Cat\left(\lvert S \rvert^{0.5}\cos\left({\theta_{S}}\right), \lvert S \rvert^{0.5}\sin\left({\theta_{S}}\right)\right)$ as target. All the models are optimized using Adam~{\cite{kingma2014adam}} with the learning rate of 5e-4. 80 epochs are conducted for network training in total, and the batch size is set to 4 at the utterance level. \textbf{The processed samples are available online, where the source code is also released.}{\footnote{https://github.com/yuguochencuc/DB-AIAT}
		
		\vspace{-3mm} 
		\section{Results and Analysis\label{Section4}}
		\label{Sec4}
		\vspace{-3mm}

		\renewcommand\arraystretch{0.95}
		\begin{table*}[t!]
			\caption{Comparison with other state-of-the-art methods including time and T-F domain methods. "$-$" denotes that the result is not provided in the original paper.}
			\setcounter{table}{0}
			\label{tbl:VB-results}
			\centering
			\small
			\scalebox{0.86}{
				\begin{tabular}{l|l|l|l|cccccc}
					\hline
					\multicolumn{1}{l|}{\textbf{Methods}}&\textbf{Year} &\multicolumn{1}{l|}{\textbf{Feature type}} &\multicolumn{1}{l|}{\textbf{Param.}}& \textbf{PESQ} & \textbf{STOI(\%)}  & \textbf{CSIG} & \textbf{CBAK} & \textbf{COVL} &\textbf{SSNR}  \\ \hline
					\multicolumn{1}{l|}{Noisy} & {\makecell[c]{--}} & {\makecell[c]{--}} & \multicolumn{1}{l|}{\makecell[c]{--}} & 1.97  & 92.1 & 3.35 & 2.44 & 2.63 & 1.68\\ \hline
					\multicolumn{9}{c}{\textbf{SOTA time and T-F Domain approaches}} \\ \hline
					\multicolumn{1}{l|}{SEGAN~{\cite{pascual2017segan}}} &2017 & \multicolumn{1}{l|}{Waveform} &\multicolumn{1}{l|}{43.2 M} & 2.16 & 92.5 & 3.48 & 2.94 & 2.80 &7.73\\ 
					\multicolumn{1}{l|}{MMSEGAN~{\cite{soni2018time}}} &2018 &\multicolumn{1}{l|}{Gammatone } &\multicolumn{1}{l|} {\makecell[c]{--}}& 2.53 & 93.0 & 3.80 & 3.12 & 3.14 & {\makecell[c]{--}} \\ 	
					\multicolumn{1}{l|}{MetricGAN~{\cite{fu2019metricgan}}} &2019 &\multicolumn{1}{l|}{Magnitude} &\multicolumn{1}{l|}{1.86 M}& 2.86 & {\makecell[c]{--}} & 3.99 & 3.18 & 3.42 & {\makecell[c]{--}}\\ 
					\multicolumn{1}{l|}{CRGAN~{\cite{zhang2020loss}}}&2020 &\multicolumn{1}{l|}{Magnitude} &\multicolumn{1}{l|}{\makecell[c]{--}} & 2.92 & 94.0 & 4.16 & 3.24 & 3.54 & {\makecell[c]{--}}\\ 
					\multicolumn{1}{l|}{DCCRN~{\cite{hu2020dccrn}}}&2020 & \multicolumn{1}{l|}{RI components} &\multicolumn{1}{l|}{3.7 M} & 2.68 & 93.7 & 3.88 & 3.18 & 3.27 & 8.62 \\ 		
					\multicolumn{1}{l|}{RDL-Net~{\cite{nikzad2020deep}}} &2020 & \multicolumn{1}{l|}{Magnitude}& \multicolumn{1}{l|}{3.91 M} & 3.02 & {\makecell[c]{93.8}}& 4.38 & 3.43 & 3.72 & {\makecell[c]{--}}\\ 
					
					\multicolumn{1}{l|}{PHASEN~{\cite{yin2020phasen}}}&2020 & \multicolumn{1}{l|}{Magnitude+Phase} &\multicolumn{1}{l|}{\makecell[c]{--}}& 2.99 & {\makecell[c]{--}} & 4.21 & 3.55 & 3.62 & 10.18 \\
					\multicolumn{1}{l|}{MHSA-SPK~{\cite{koizumi2020speech}}}&2020 & \multicolumn{1}{l|}{Waveform} &\multicolumn{1}{l|}{\makecell[c]{--}}& 2.99 & {\makecell[c]{--}} & 4.15 & 3.42 & 3.53 & {\makecell[c]{--}} \\ 
					\multicolumn{1}{l|}{T-GSA~{\cite{kim2020t}}} &2020 & \multicolumn{1}{l|}{RI components}&\multicolumn{1}{l|}{\makecell[c]{--}} & 3.06 & 93.7 & 4.18 & 3.59 & 3.62 &10.78 \\ 
					\multicolumn{1}{l|}{TSTNN~{\cite{wang2021tstnn}}} &2021 & \multicolumn{1}{l|}{Waveform} &\multicolumn{1}{l|}{0.92 M}& 2.96 & 95.0 & 4.17 & 3.53 & 3.49 & 9.70\\ 
					\multicolumn{1}{l|}{DEMUCS ~{\cite{defossez2020real}}}&2021 & \multicolumn{1}{l|}{Waveform} &\multicolumn{1}{l|}{128 M}& 3.07 & 95.0 & 4.31 & 3.40 & 3.63 & {\makecell[c]{--}} \\ 
					\multicolumn{1}{l|}{GaGNet~{\cite{li2021glance}}} &2021& \multicolumn{1}{l|}{Magnitude+RI} &\multicolumn{1}{l|}{5.94 M}& 2.94 & 94.7 & 4.26 & 3.45 & 3.59 & 9.24\\ 
					\multicolumn{1}{l|}{MetricGAN+~{\cite{fu2021metricgan+}}} &2021 &\multicolumn{1}{l|}{Magnitude} &\multicolumn{1}{l|}{\makecell[c]{--}}& 3.15 & {\makecell[c]{--}} & 4.14 & 3.16 & 3.64 & {\makecell[c]{--}}\\ 
					\multicolumn{1}{l|}{SE-Conformer~{\cite{eesung2021se}}} &2021 &\multicolumn{1}{l|}{Waveform} &\multicolumn{1}{l|}{\makecell[c]{--}}& 3.13 & 95.0 & 4.45 & 3.55 & 3.82 & {\makecell[c]{--}}\\
					\hline
					
					\multicolumn{9}{c}{\textbf{Proposed approaches}} \\ \hline
					\multicolumn{1}{l|}{MMB-AIAT} &2021&  \multicolumn{1}{l|}{Magnitude} &\multicolumn{1}{l|}{0.90 M} & 3.11 & 94.9 & 4.45 & 3.60 & 3.79 & 9.74\\ 
					\multicolumn{1}{l|}{CRB-AIAT} &2021&  \multicolumn{1}{l|}{RI components} &\multicolumn{1}{l|}{1.17 M} & 3.15 & 94.7 & 4.48 & 3.54 & 3.81 & 8.81\\ 
					\multicolumn{1}{l|}{DB-AIAT} &2021 &  \multicolumn{1}{l|}{Magnitude+RI} &\multicolumn{1}{l|}{2.81 M} & \textbf{3.31} & \textbf{95.6} & \textbf{4.61} & \textbf{3.75} & \textbf{3.96} & \textbf{10.79}\\ \hline
				\end{tabular}
			}
			\vspace{-6mm}
		\end{table*}

		\renewcommand\arraystretch{1.3}
		\begin{table}[t!]
			\setcounter{table}{1}
			\caption{Ablation study $\emph{w.r.t.}$ dual-branch strategy and attention-in-attention transformer structure.}
			\centering
			\scalebox{0.72}{
				\begin{tabular}{l|cc|ccccc}
					\hline
					\multirow{2}*{Models}
					&\multirow{2}*{\shortstack{ATAB\\/AFAB}}  &\multirow{2}*{AHA} &\multirow{2}*{PESQ}   &\multirow{2}*{STOI(\%)}  &\multirow{2}*{CSIG} &\multirow{2}*{CBAK} &\multirow{2}*{COVL}   \\ & & & & & &  &  \\
					\hline
					Unprocessed &\makecell[c]{--} & \makecell[c]{--}  &1.97 &92.1 &3.35 &2.44 &2.63 \\ 
					\hline
					\multicolumn{8}{c}{\textbf{Single-Branch approaches}} \\ 
					
					\hline
					MMB-ATFAT   &\Checkmark/\Checkmark &\XSolidBrush &3.05 &94.6 &4.37 &3.53 & 3.71  \\ 
					\hline
					MMB-AIAT  &\Checkmark/\Checkmark &\Checkmark  &3.11 &94.9 &4.45 &3.60 &3.79  \\
					\hline
					CRB-ATFAT  &\Checkmark/\Checkmark &\XSolidBrush  &3.07 &94.5 &4.40 &3.52 & 3.72 \\     
					\hline
					CRB-AIAT   &\Checkmark/\Checkmark &\Checkmark  &3.15 &94.7 & 4.48 & 3.54 & 3.81 \\  
					
					\hline
					\multicolumn{8}{c}{\textbf{Dual-Branch approaches}} \\ 	
					\hline
					DB-ATAT    &\Checkmark/\XSolidBrush &\XSolidBrush &2.82 &94.2 &4.17 &3.29 & 3.47  \\
					\hline
					DB-AFAT   &\XSolidBrush/\Checkmark &\XSolidBrush &2.93 &94.4 &4.28 &3.31 & 3.63  \\  
					\hline
					DB-ATFAT   &\Checkmark/\Checkmark &\XSolidBrush &3.18 &95.0 &4.50 &3.68 & 3.86  \\
					\hline
					DB-AIAT  &\Checkmark/\Checkmark  &\Checkmark &3.31 &95.6 &4.61 &3.75 & 3.96  \\  
					\hline
				\end{tabular}
			}
			\label{tbl:ablation-study}
			\vspace{-0.55cm}
		\end{table}
		
		We use the following objective metrics to evaluate speech enhancement performance: the perceptual evaluation of speech quality (PESQ)~{\cite{rix2001perceptual}}, short-time objective intelligibility (STOI)~{\cite{taal2010short}}, segmental signal-to-noise ratio (SSNR), the mean opinion score (MOS) prediction of the speech signal distortion (CSIG)~{\cite{hu2007evaluation}}, the MOS prediction of the intrusiveness of background noise (CBAK) and the MOS prediction of the overall effect (COVL)~{\cite{hu2007evaluation}} to evaluate speech enhancement performance. Higher values of all metrics indicate better performance.
		\vspace{-4mm}  
		\subsection{Comparison with previous advanced baselines\label{Section42}}
		\vspace{-2mm}  
		We first compare the objective performance of the proposed methods with other state-of-the-art (SOTA) baselines, whose results are presented in Table~{\ref{tbl:VB-results}}. The baselines include five time-domain methods (\emph{e.g.}, SEGAN~{\cite{pascual2017segan}}, MHSA-SPK~{\cite{koizumi2020speech}}, TSTNN~{\cite{wang2021tstnn}} and DEMUCS~{\cite{defossez2020real}}) and nine T-F domain methods(\emph{e.g.}, CRGAN~{\cite{zhang2020loss}}, PHASEN~{\cite{yin2020phasen}}, T-GSA~{\cite{kim2020t}}, GaGNet~{\cite{li2021glance}} and MetricGAN+~{\cite{fu2021metricgan+}}). One can have the following observations. 
		First, when only either the magnitude masking branch (MMB-AIAT) or the complex refining branch (CRB-AIAT) is adopted, the proposed method dramatically achieves competitive performance compared with the most advanced single-branch baselines. 
		For example, going from CRGAN to MMB-AIAT, average 0.19, 0.9\%, 0.29, 0.36 and 0.25 improvements are achieved in terms of PESQ, STOI, CSIG, CBAK and COVL, respectively. Similarly, CRB-AIAT provides average 0.47 PESQ, 1.0\% STOI, 0.60 CSIG, 0.36 CBAK and 0.54 COVL improvements over DCCRN. This verifies the effectiveness of the proposed attention-in-attention transformer in improving speech quality.
		Second, by simultaneously adopting two branches in parallel, DB-AIAT consistently surpasses existing SOTA time and T-F domain methods in terms of most metrics. For example, DB-AIAT provides average 0.24, 0.6\%, 0.30, 0.35 and 0.33 improvements over DEMUCS in terms of PESQ, STOI, CSIG, CBAK and COVL, respectively. Third, we also provide the comparison of the number of parameters between our methods with some SOTA methods, as presented in Table 2. One can find that DB-AIAT has a relatively lower parameter burden (2.81 M) compared with other SOTA peers. 
		\vspace{-6mm}
		\subsection{The effects of Dual-branch strategy and AIAT structure\label{Section41}}
		\vspace{-1mm}
		We then investigate the effects of the proposed dual-branch strategy and AIAT structure, as shown in Table~{\ref{tbl:ablation-study}}. From the results, one can have the following observations. 
		First, when adopting the single-branch topology, \emph{i.e.}, MMB-AIAT and CRB-AIAT, we can find that CRB-AIAT yields better performance in PESQ, STOI, CBAK and COVL than MMB-AIAT, while MMB-AIAT achieves a higher score in CBAK. This indicates that MMB can better eliminate noise and provide higher STOI scores, while CRB conducts better speech overall quality. Second, when both CRB and MMB are employed in parallel, DB-AIAT yields significant improvements in terms of all metrics than the single-branch methods. This verifies that merging two branches can collaboratively facilitate the spectrum recovery from the complementary perspective. Third, we investigate the effectiveness of our proposed attention-in-attention transformer (AIAT) model. When combining ATAB and AFAB to capture both temporal-frequency dependency, DB-ATFAT consistently surpasses DB-ATAT and DB-AFAT, where DB-ATAT and DB-AFAT only capture long-term dependencies along time dimension and that along frequency dimension, respectively. For example, DB-ATFAT provides average 0.36, 0.8\%, 0.33, 0.39 and 0.39 improvements than DB-ATAT in PESQ, STOI, CSIG, CBAK and COVL, respectively. Besides, going from DB-ATFAT to DB-AIAT by adding the AHA module, relatively better performance can be achieved, which reveals the effectiveness of the proposed AIAT in further improving speech quality.

		\vspace{-3mm}
		\section{Conclusions\label{Section5}}
		\label{Sec5}
		\vspace{-3mm}
		
		In this paper, we propose a dual-branch transformer-based method to collaboratively recover the clean complex spectrum from the complementary perspective. To be specific, we employ a magnitude masking branch (MMB) to coarsely estimate the magnitude spectrum of clean speech, and the residual spectral details are derived in parallel by a complex refining branch (CRB). With interconnection between each branch, MMB aims at estimating the overall magnitude of the clean spectrum, while CRB can simultaneously compensate for some missing complex spectral details and restore phase information. Each branch incorporates an attention-in-attention transformer (AIAT) module between a densely encoder-decoder architecture for temporal sequence modeling, which effectively strengthens long-range temporal-frequency dependencies and aggregates global hierarchical contextual information. Comprehensive experiments on the public dataset demonstrate that the proposed method achieves state-of-the-art performance over previous competitive systems.
		
		\vspace{-3mm}

		\bibliographystyle{IEEEbib}
		\begin{spacing}{0.9} 
			\bibliography{myrefs} 
		\end{spacing} 
		
	\end{document}